
\magnification=1200
\hsize 13true cm \hoffset=0.5true cm
\vsize 20true cm
\baselineskip=11pt

\def\laq{\raise 0.4ex\hbox{$<$}\kern -0.8em\lower 0.62
ex\hbox{$\sim$}}
\def\gaq{\raise 0.4ex\hbox{$>$}\kern -0.7em\lower
0.62 ex\hbox{$\sim$}}

\font\small=cmr8 scaled \magstep0
\font\grande=cmr10 scaled \magstep4
\font\medio=cmr10 scaled \magstep2
\outer\def\beginsection#1\par{\medbreak\bigskip
      \message{#1}\leftline{\bf#1}\nobreak\medskip\vskip-
\parskip
      \noindent}

\def \pa {\partial}
\def \ra {\rightarrow}

\def \fb {\overline \phi}
\def \rb {\overline \rho}

\def \ti {\tilde}
\def \la {\lambda}

\def \Da {\Delta}
\def \b {\beta}
\def \a {\alpha}
\def \ap {\alpha^\prime}

\def \ga {\gamma}
\def \sg {\sigma}
\def \da {\delta}

\def \r {\rho}

\def \e {\eta}
\def \es {\eta_s}
\def \om {\omega}
\def \Om {\Omega}
\def \noi {\noindent}

\def\sqr#1#2{{\vcenter{\hrule height.#2pt\hbox{\vrule
width.#2pt
height#1pt \kern#1pt\vrule width.#2pt}\hrule height.#2pt}}}

\def\lsim{\mathrel{\rlap{\lower4pt\hbox{\hskip1pt$\sim$}}
    \raise1pt\hbox{$<$}}}         
\def\gsim{\mathrel{\rlap{\lower4pt\hbox{\hskip1pt$\sim$}}
    \raise1pt\hbox{$>$}}}         

\nopagenumbers
\line{\hfil DFTT-55/95}
\line{\hfil September 1995}
\line{\hfil hep-th/9509127}

\vskip 1.3 true cm
\centerline {\grande Status of String Cosmology:}
\vskip 0.4 true cm
\centerline{{\grande Phenomenological Aspects}
\footnote{*}{\small Lecture given at the 4th Course of the International
School of Astrophysics ``D. Chalonge" on
{\it {\bf String gravity and physics at the Planck scale}}
(Erice, 8-19 September 1995).
To appear in the {\bf Proceedings}
(Kluwer Acad. Pub., Dordrecht), ed. by N. Sanchez.}}

\vskip 1.3true cm
\centerline{M. Gasperini}
\bigskip
\centerline{\it Dipartimento di Fisica Teorica, Universit\`a di
Torino,}
\centerline{\it Via P.Giuria 1, 10125 Turin, Italy,}
\centerline{and}
\centerline{\it INFN, Sezione di Torino, Turin, Italy}

\vskip 1.3 true cm
\centerline{\medio Abstract}
\bigskip
\noindent
I report recent studies on
the evolution of perturbations in the context of the
``pre-big-bang" scenario typical of string cosmology, with emphasis on
the formation of a stochastic
background of relic photons and gravitons, and
its possible direct/indirect observable consequences. I also discuss the
possible generation of a thermal microwave background by using, as
example, a simple gravi-axio-dilaton model whose classical evolution
connects smoothly inflationary expansion to decelerated contraction. By
including the quantum back-reaction of the produced radiation the model
eventually approaches
the standard radiation-dominated (constant dilaton) regime.

\vfill\eject

\footline={\hss\rm\folio\hss}
\pageno=1

\baselineskip=11pt

\topinsert
\vskip 1.5 cm
\endinsert

\noi
{\bf STATUS OF STRING COSMOLOGY:}

\noi
{\bf PHENOMENOLOGICAL ASPECTS}
\vskip 1.5 cm
\hskip 1 cm {M. GASPERINI}

\hskip 1cm{\it Dipartimento di Fisica Teorica,
Universit\`a di Torino,}

\hskip 1 cm{\it Via P.Giuria 1, 10125 Turin, Italy}
\vskip 1.5 cm
\noi
{\bf 1. Introduction}
\bigskip
\noi
Inspired by the basic ideas of string theory, we have recently started
the investigation of a cosmological scenario in which the standard
big-bang singularity is smoothed out and is replaced by a phase of
maximal (finite) curvature. Such a phase is preceeded in time by a
``pre-big-bang" epoch [1-3], which has (approximately) specular
properties with respect to the present phase of standard decelerated
evolution.

This scenario was originally motivated by the solutions of the string
equations of motion in curved backgrounds [4], and by the duality
symmetries of the string effective action [5-7] (see for instance [8]
for a more detailed introduction). Its basic ingredients are an initial
weak coupling, perturbative regime characterized by a very small
background curvature (in string units), and a final transition to the
radiation era controlled by the dilaton dynamics and by the two basic
parameters (string length, string coupling) of string theory. The most
revolutionary aspect of this scenario (with respect to the conventional,
even inflationary, picture)
is probably the fact that the early, pre-Planckian universe can be
consistently described in terms of a semiclassical low energy effective
action, with the vacuum as the most natural initial conditions for the
quantum fluctuations of all the background fields (more details on this
scenario can be found in Gabriele Veneziano's contribution to these
proceedings [9]).

According to the above scenario, the pre-big-bang epoch is characterized
by an accelerated (i.e. inflationary) evolution. In all
inflationary models, the most direct (and probably most spectacular)
phenomenological predictions is the parametric amplification of
perturbations [10], and the corresponding generation of primordial
spectra, directly from the quantum fluctuations of the background
fields. In a string cosmology context such an effect is even more
spectacular, as the growth of the curvature scale during the
pre-big-bang phase is associated with a perturbation spectrum which
grows with frequency [11,12]; moreover, the growth of the curvature may
also force the comoving perturbation amplitude to grow (instead of being
frozen) outside the horizon, as first noted in [2]. This leads to a more
efficient
amplification of perturbations, but the amplitude could grow
too much, during the pre-big-bang phase, so as to prevent us from
applying the standard linearized formalism.

In view of this aspect, the aim of this lecture is twofold. On one
hand I will discuss how, in some case, this anomalous growth
can be gauged away, so that perturbations can consistently
linearized in an appropriate frame. On the other hand I will
show that, even if the growth is physical, and we have to
restrict ourself to a reduced portion of parameter space in
order to apply a linearized approach, such enhanced
amplification is nevertheless rich of interesting
phenomenological consequences.

I will discuss in particular the following points: {\it i)}
the growing mode of scalar metric perturbations in a
dilaton-driven background, which appears in the standard
longitudinal gauge and which seems to complicate the
computation of the spectrum [13];
{\it ii)} the production
of a relic gravity wave background with a spectrum strongly
enhanced in the high frequency sector, and its possible
observation by large interferometric detectors [14] (such as
LIGO and VIRGO); {\it iii)} the amplification of electromagnetic
perturbations due to their direct coupling to the dilaton
background, and the generation of primordial ``seeds" for the
galactic and extragalactic magnetic field [15]; {\it iv)} the
generation of the large scale CMB anisotropy directly from the
vacuum fluctuations of the electromagnetic field [16]. Finally, I will
present some speculations about a possible geometric origin of the CMB
radiation itself [17]. This lecture is an extended version of previous
lectures given at the Observatory of Paris [18] and at the Gaeta
Workshop on the
``Very Early Universe" (Gaeta, August 1995, unpublished).
The main results reported here are based on recent work done in
collaboration with R. Brustein, M. Giovannini, V. Mukhanov and G.
Veneziano.

Throughout this lecture the evolution of perturbations will be
discussed in a type of background  which I will call, for
short, ``string cosmology background". At low energy such
background represents a solution [3,5] of the gravi-dilaton effective
action $S$, at
tree-level in the string coupling $e^{\phi/2}$, and to
zeroth order in the inverse string tension $\ap$,
$$
S=
-\int d^{d+1}x \sqrt{|g|} e^{-\phi}\left(R+\pa_\mu\phi\pa^\mu
\phi\right) \eqno(1.1)
$$
(with the possible addition of string matter sources). The background
describes the accelerated evolution from the string perturbative
vacuum, with flat metric and vanishing dilaton coupling
($\phi=-\infty$), towards a phase driven by the kinetic energy
of the dilaton field ($H^2\sim \dot\phi^2$), with negligible
contribution from the dilaton potential (and other matter sources).
In this initial phase the
curvature scale $H^2$ and the dilaton coupling $e^\phi$ are
both growing at a rate uniquely determined by
the action (1.1), and the possible presence of matter, in the form of a
perfect gas of non-mutually interacting classical strings, is eventually
diluted [3,19].

The background evolution can be consistently described in terms of the
action (1.1), however, only up to the time $t=t_s$ when the
curvature reaches the string scale, namely when $H\simeq
H_s= (\ap)^{-1/2}\equiv \la_s^{-1}$. At that time all higher orders in
$\ap$ (i.e. all higher-derivative corrections to the string effective
action) become important,
and the background enters a truly ``stringy" phase, whose
kinematic details cannot be predicted on the ground of the
previous simple action. The presence of this high-curvature
phase cannot be avoided, as it is required [20] to stop the
growth of the curvature, to freeze out the dilaton, and to
arrange a smooth transition (at $t=t_1$) to the standard
radiation-dominated evolution (where $\phi=$const).

In previous works (see for instance [3,8]) we assumed that the
time scales $t_s$ and $t_1$ (marking respectively the
beginning of the string and of the radiation era) were of the
same order, and we computed the perturbation spectrum in
the sudden approximation, by matching directly the radiation
era to the dilaton-driven phase. Here I will consider a more
general situation in which the duration of the string era
($t_1/t_s$) is left completely arbitrary, and I will discuss its
effects on the perturbation spectrum.

During the pre-big-bang epoch, from
the flat and cold initial state to the highly
curved (and strongly coupled) final regime,
the background evolution is accelerated, and
can be invariantly characterized, from a kinematic
point of view, as a phase of shrinking event horizons [1-3].
If we parameterize the pre-big-bang scale factor
(in cosmic time) as
$$
a(t)\sim (-t)^\b, \,\,\,\,\,\,\,\,\,\,\, -\infty < t < 0 \eqno(1.2)
$$
then the condition for the existence of shrinking event horizons
$$
\int_t^0 {dt'\over a(t')} < \infty \eqno(1.3)
$$
is simply  $\b<1$. As a consequence, there are two
physically distinct classes of backgrounds in
which the event horizon is shrinking.

If $\b<0$ we have a metric describing a phase of
accelerated expansion and growing curvature,
$$
\dot a >0 , \,\,\,\,\,\,\, \ddot a >0 , \,\,\,\,\,\,\, \dot H>0
\eqno(1.4)
$$
of the type of pole-inflation [21], also called super-inflation
($H=\dot a/a$, and a dot denotes differentiation with respect to
the cosmic time $t$). If $0<\b<1$ we have instead a metric
describing accelerated contraction and growing curvature
scale,
$$
\dot a <0 , \,\,\,\,\,\,\, \ddot a <0 , \,\,\,\,\,\,\, \dot H<0
\eqno(1.5)
$$
The first type of metric provides a representation of the
pre-big-bang scenario in the String (or Brans-Dicke) frame, in
which test strings move along geodesic surfaces. The second in
the Einstein frame, in which the gravi-dilaton action is
diagonalized in the standard canonical form (see for instance [2,3]).

In both types of backgrounds the computation of the metric
perturbation spectrum may become problematic, but the best
frame to illustrate the difficulties is probably the Einstein
frame, where the metric is contracting. It should be recalled,
in this context, that the tensor perturbation spectrum for
contracting backgrounds was first given in [22],
but the possible occurrence of problems, due to a
growing solution of the perturbation equations, was pointed
out only much later [23], in the context of dynamical
dimensional reduction. The problem, however, was left
unsolved.
\vskip 1.5 cm
\noi
{\bf 2. The ``growing mode" problem}
\bigskip
\noi
Consider the evolution of tensor metric perturbations, $\da
g_{\mu\nu}=a^2h_{\mu\nu}$, in a $(3+1)$-dimensional
conformally flat background, parameterized in conformal time
($\eta=\int dt/a$) by the scale factor
$$
a(\eta)\sim (-\eta)^\a, \,\,\,\,\,\,\,\,\,\,\, -\infty < \eta < 0
\eqno(2.1)
 $$
Define the correctly normalized variables $u_{\mu\nu}=aM_ph_{\mu\nu}$
($M_p$ is the Planck mass),  whose
Fourier components obey
canonical commutation relations, $[u_k,\dot u_{k'}]= i \da_{k,k'}$.
The modes $u_k$
satisfy, for each of the two physical (transverse traceless)
polarizations, the well known perturbation equation [10]
$$
u_k''+(k^2-{a''\over a})u_k=0\eqno(2.2)
$$
(a prime denotes differentiation with respect to $\eta$). In a
string cosmology background the horizon is shrinking, so that
all comoving length scales $k^{-1}$ are ``pushed out"of the
horizon. For a mode $k$ whose wavelength is larger than the
horizon size (i.e. for $|k\eta|<<1$), we have then the general (to
leading order) asymptotic
solution
$$
h_k= {u_k\over a M_p}=A_k+B_k|\eta|^{1-2\a} , ~~~~~~~~
\,\,\,\,\, \eta \ra0_- \eqno(2.3)
$$
where $A_k$ and $B_k$ are integration constants.

The asymptotic behavior of the perturbation is thus
determined by $\a$. If $\a <1/2$ the perturbation tends to stay
constant outside the horizon, and the typical amplitude
$|\da_h|$ at the scale $k^{-1}$, for modes normalized to an
initial vacuum fluctuation spectrum,
$$
\lim_{\eta \ra -\infty} u_k \sim {1\over \sqrt k} e^{-ik\eta}
\eqno(2.4)
$$
can be given as usual [24] in terms of the Hubble factor at
horizon crossing ($k\eta \sim 1$)
$$
|\da_h|=k^{3/2}|h_k|\simeq \left( H\over
M_p\right)_{HC}\eqno(2.5)
$$
In this case the amplitude is always smaller than one provided
the curvature is smaller than Planckian. This case includes in
particular $\a<0$, namely all backgrounds describing,
according to eq. (2.1),
accelerated inflationary expansion.

If, on the contrary, $\a>1/2$, then the second term is the dominant
one in the solution (2.3), the perturbation amplitude tends to
grow outside the horizon,
$$
|\da_h|=k^{3/2}|h_k|\simeq \left (H\over
M_p\right)_{HC}|k\eta|^{1-2\a} , ~~~~~~~~
\,\,\,\,\,\, \eta \ra 0_-
\eqno(2.6)
$$
and may become larger than one, thus breaking the validity of
the perturbative approach. Otherwise stated: the energy
density (in critical units) stored in the mode $k$, i.e.
$\Om(k)=d(\r/\r_c)/d\ln k$, may become larger than one, in
contrast with the hypothesis of negligible back-reaction of
the perturbations on the initial metric.

One might think that this problem - due to the dominance of
the second term in eq. (2.3) - appears in the Einstein frame
because of the contraction (which
corresponds to $\a>0$), but disappears in the String frame
where the metric is expanding. Unfortunately this is not true
because, in the String frame, the different metric background
is compensated by a different perturbation equation, in such a
way that the perturbation spectrum remains exactly the same
[2,3].

This important property of perturbations can be easily
illustrated by taking, as a simple example, an isotropic solution
of the $(d+1)$-dimensional gravi-dilaton equations [2,3],
obtained from the action (1.1) supplemented by a perfect gas
of long, stretched strings as sources (with equation of state
$p=-\r/d$).

In the Einstein frame the solution describes a metric background which
is contracting for $\eta \ra 0_-$,
$$
a=\left(-\eta\right)^{2(d+1)/(d-1)(3+d)} , \,\,\,\,\,\,
\phi = -{4d\over 3+d}\sqrt{2\over d-1} \ln (-\eta) \eqno (2.7)
$$
and the tensor perturbation equation
$$
h_k'' +(d-1){a'\over a}h_k' +k^2h_k=0 \eqno(2.8)
$$
has an asymptotic solution (for $|k\eta|<<1$) which grows,
according to eq. (2.3), as
$$
\lim_{\eta \ra 0_-} h_k \sim |\eta |^{(1-d)/(d+3)} \eqno (2.9)
$$
In the String frame the metric is expanding,
$$
\ti a=\left(-\eta\right)^{-2/(3+d)} , \,\,\,\,\,\,\,\,\,
\ti \phi = -{4d\over 3+d} \ln (-\eta) \eqno (2.10)
$$
but the perturbation is also coupled to the time-variation of
the dilaton background [12],
$$
h_k'' +\left[(d-1){\ti a'\over \ti a}- \ti \phi'\right]h_k'
+k^2h_k=0 \eqno(2.11)
$$
As a consequence, the explicit form of the perturbation
equation is exactly the same as before,
$$
h_k'' +{2(d+1)\over d+3} {h_k'\over \eta} +k^2h_k=0 \eqno(2.12)
$$
so that the solution is still growing, asymptotically, with the
same power as in eq. (2.9).

It may be noted that in the String frame the growth of
perturbations outside the horizon is due to the joint
contribution of the metric and of the dilaton background to the
``pump" field responsible for the parametric amplification
process [25]. Such an effect is thus to be expected in generic
scalar-tensor backgrounds, as noted also in [26]. The particular
example chosen above is not much relevant, however, for a
realistic scenario in which the phase of pre-big-bang inflation
is long enough to solve the standard cosmological problems. In
that case, in fact, all scales which are inside our present
horizon crossed the horizon (for the first time) during the
dilaton-driven phase or during the final string phase, in any
case when the contribution of matter sources was negligible
[3,8].

We shall thus consider, as a more significant (from a
phenomenological point of view) background, the vacuum,
dilaton-driven solution of the action (1.1), which in the Einstein
frame (and in $d=3$) can be explicitly written as
$$
a=(-\eta)^{1/2} , \,\,\,\,\,\,\,\,\,\,\,\, \phi=-\sqrt 3 \ln (-\eta) ,
\,\,\,\,\,\,\,\,\,\,\,\,  -\infty < \eta <0 \eqno(2.13)
$$
In such a background one finds that the growth of tensor
perturbations is simply logarithmic [13],
$$
|\da_h(\eta)|\simeq \left |H\over
M_p\right|_{HC}\ln|k\eta| \simeq {H_s\over
M_p}|k\eta_s|^{3/2}\ln|k\eta| ,~~~ \,\,
|k\eta_s|<1 , \,\, |\eta|>|\eta_s|
\eqno(2.14)
$$
so that it can be easily kept under control, provided the
curvature scale $H_s\sim (a_s\eta_s)^{-1}$ at the end of the
dilaton phase is bounded.

The problem, however, is with scalar perturbations, described
in the longitudinal gauge by the variable $\psi$ such that [24]
$$
(g_{\mu\nu}+\da g_{\mu\nu})dx^\mu dx^\nu=
a^2(1+2\psi)d\eta^2 -a^2(1-2\psi) (dx_i)^2\eqno(2.15)
$$
The canonical variable $v$ associated to $\psi$ is defined (for
each mode $k$) by [24]
$$
\psi_k= -{\phi'\over 4k^2M_p} \left(v_k\over a\right)'
\eqno(2.16)
$$
and satisfies a perturbation equation
$$
v_k''+(k^2-{a''\over a})v_k=0\eqno(2.17)
$$
which is identical to eq. (2.2) for the tensor canonical
variable, with asymptotic solution
$$
{v_k\over a}\simeq {1\over \sqrt k} {\ln |k\eta|\over a_{HC}} ,
\,\,\,\,\,\,\,\,\,\,\,~~~~~~~~~~~~~~ |k\eta|<<1 \eqno(2.18)
$$
Because of the different relation between canonical variable
and metric perturbation, however, it turns out that the
amplitude of longitudinal perturbations, normalized to an initial
vacuum fluctuation spectrum,
$$
\lim_{\eta \ra -\infty} v_k \sim {1\over \sqrt k} e^{-ik\eta}
\eqno(2.19)
$$
grows, asymptotically, like $\eta^{-2}$. We have in fact, from
(2.16),
$$
|\da_\psi(\eta)|= k^{3/2}|\psi_k|\simeq \left |H\over
M_p\right||k\eta| ^{-1/2} \simeq
 \left |H\over
M_p\right|_{HC}|k\eta| ^{-2}\simeq
$$
$$
\simeq \left (H_s\over
M_p\right){|k\eta_s|^{3/2}\over |k\eta|^2} \sim {1\over
\eta^2} , \,\,\,\,\,\,~~~~~~ \eta \ra 0_-
\eqno(2.20)
$$

This growth, as we have seen, cannot be eliminated by passing
to the String frame. Neither can be eliminated in a background
with a higher number of dimensions. In fact, in $d>3$, the
isotropic solution (2.13) is generalized as [3]
$$
a=(-\eta)^{1/(d-1)} , \,\,\,\,\,\,\,\,\,\,\,\, \phi=-\sqrt
{2d(d-1)} \ln a ,
 \,\,\,\,\,\,\,\,\,\,\,\,  -\infty < \eta <0 \eqno(2.21)
$$
and the scalar perturbation equation in the longitudinal gauge
[3]
$$
\psi_k'' +3(d-1){a'\over a}\psi_k' +k^2\psi_k=0 \eqno(2.22)
$$
has the generalized asymptotic solution
$$
\psi_k= A_k +B_k\eta a^{-3(d-1)} \eqno(2.23)
$$
By inserting the new metric (2.21) one thus finds the same
growing time-behavior, $\psi_k\sim \eta^{-2}$, exactly as
before. The same growth of $\psi_k$ is also found in
anisotropic, higher-dimensional, dilaton-dominated
backgrounds [13].

Because of the growing mode there is always (at any given
time $\eta$) a low frequency band for which $|\da_\psi
(\eta)|>1$. In $d=3$, in particular, such band is defined [from
eq.(2.20)] as $k<\eta^{-1}(H/M_p)^2$. For such modes the
linear approximation breaks down in the longitudinal gauge,
and a full non-linear treatment would seem to be required in
order to compute the spectrum. In spite of this conclusion, a
linear description of scalar perturbations may remain possible
provided we choose a different gauge, more appropriate to
linearization than the longitudinal one.

A first signal that a perturbative expansion around a
homogeneous background can be consistently truncated at the
linear level, comes from an application of the ``fluid flow"
approach [27,28] to the perturbations of a scalar-tensor
background. In this approach, the evolution of density and
curvature inhomogeneities is described in terms of two
covariant scalar variables, $\Da$ and $C$, which are gauge
invariant to all orders [29]. They are defined in terms of the
momentum density of the scalar field, $\nabla \phi$, of the
spatial curvature, $^{(3)}R$, and of their derivatives. By
expanding around our homogeneous dilaton-driven background
(2.13) one finds [13] for such variables, in the linear
approximation,  the asymptotic solution
($|k\eta|<<1$),
$$
\Da_k = const , \,\,\,\,\,\,\,~~~~~~~~~~
c_k= const +A_k \ln |k\eta| \eqno(2.24)
$$
This solution shows that such variables
tend to stay constant outside the horizon,
with at most a logarithmic variation (like in the tensor case),
which is not dangerous.

As a consequence, the amplitude of density and curvature
fluctuations can be consistently computed in the linear
approximation (for all modes) in terms of $\Da$ and $C$, and
their spectral distribution (normalized to an initial vacuum
spectrum) turns out to be exactly the same as the tensor
distribution (2.14), which is bounded.

What is important, moreover, is the fact that such a spectral
distribution could also be obtained directly from the
asymptotic solution of the scalar perturbation equations in the
longitudinal gauge [13],
$$
\psi_k = c_1 \ln |k\eta| + {c_2\over \eta^2} \eqno (2.25)
$$
simply by neglecting the growing mode contribution (i.e.
setting $c_2=0$). This may suggests that such growing mode
has no direct physical meaning, and that it should be possible
to get rid of it through an appropriate coordinate choice.

A good candidate to do the job is what we have called [13]
``off-diagonal" gauge,
$$
(g_{\mu\nu}+\da g_{\mu\nu})dx^\mu dx^\nu=
a^2\left[(1+2\varphi)d\eta^2 -(dx_i)^2-2\pa_iB dx^i d\eta\right]
\eqno(2.26)
$$
which represents a complete choice of coordinates, with no
residual degrees of freedom, just like the longitudinal gauge.
In this gauge there are two variables for scalar perturbations,
$\varphi$ and $B$, and their asymptotic solution is, in the linear
approximation,
$$
\varphi_k= c_1 \ln |k\eta| \sim \psi_k , \,\,\,\,\,\,~~~~
B_k ={c_2\over \eta} \sim \eta \psi_k ,  \,\,\,\,\,\,~~~~
(\pa B)_k \sim |k\eta| \psi_k \eqno(2.27)
$$
($c_1$ and $c_2$ are integration constants). The growing mode
is thus completely gauged away for homogeneous
perturbations (for which $\pa_i B=0$). It is still present for
non-homogeneous perturbations in the off-diagonal part of
the metric, but it is  ``gauged down" by the factor $k\eta$
which is very small, asymptotically.

Fortunately this is enough for the validity of the linear
approximation, as the amplitude of the off-diagonal
perturbation, in this gauge, outside the horizon,
$$
|\da_B| \simeq |k\eta||\da _\psi|
 \simeq \left (H_s\over
M_p\right)|k\eta_s|^{1/2}\left|\eta_s\over \eta\right|
\eqno(2.28)
$$
stays smaller than one for all modes $k<|\eta_s|^{-1}$, and for
the whole duration of the dilaton-driven phase, $|\e|>|\es|$.
We have explicitly checked that quadratic corrections are
smaller than the linear terms in the perturbation equations,
but a full second order computation requires a further
coordinate transformation [13]. The higher order problem is very
interesting in itself, but a complete discussion of the problem is
outside the scope of this lecture. Having established that the
vacuum fluctuations of the metric background, amplified by a
phase of dilaton-driven evolution, can be consistently described
(even in the scalar case) as small corrections of the
homogeneous background solution, let me discuss instead some
phenomenological consequence of such amplification. Scalar
perturbations and dilaton production were discussed in
[3,8,30]. Here I will concentrate, first of all, on graviton
production.

\vfill\eject
\noi
{\bf 3. The graviton spectrum from dilaton-driven
inflation}
\bigskip
\noi
Consider the amplification of tensor metric perturbation in a
generic string cosmology background, of the type of that
described in Sect. 1. Their present spectral energy distribution,
$\Om (\om, t_0)$, can be computed in terms of the Bogoliubov
coefficient determining their amplification (see Sect. 5 below),
or simply by following the evolution of the typical amplitude
$|\da_h|$ from the time of horizon crossing down to the
present time $t_0$. For modes crossing the horizon in the
inflationary dilaton-driven phase ($t<t_s$), and
re-entering the horizon in the decelerated radiation era
($t>t_1$), one easily finds, from eq. (2.14),
$$
\Om (\om, t)\equiv {\om\over \r_c}{d\r \over d\om} \simeq
 A \Om_\ga \left(H_s\over M_p\right)^2
\left(\om\over \om_s\right)^3 \ln^2 \left(\om\over
\om_s\right) ,~~~t>t_1,~~~  \om <\om_s \eqno(3.1)
$$

Here $\om=k/a$ is the red-shifted proper frequency for the mode
$k$ at time $t$, $\r_c=M_p^2 H^2 $ is the critical energy
density, $\Om_\ga$ = $(H_1/H)^2(a_1/a)^4$ = $\r_\ga /\r_c$ is the
radiation energy density in critical units, and $\om_s=
H_sa_s/a$ is the maximum frequency amplified during the
dilaton-driven phase. Finally, $A$ is a possible amplification
factor due to the subsequent string phase ($t_s<t<t_1$), in
case that the perturbation amplitude
grows outside the horizon (instead
of being constant) during such phase. This additional
amplification does not modify however the slope of the
spectrum, as we are considering modes that crossed the
horizon before the beginning of the string phase (see Sect. 5).

An important property of the spectrum (3.1) is the universality
of the slope $\om^3$ with respect to the total number $d$ of
spatial dimensions, and with respect to
their possible anisotropy. Actually, the
spectrum is also duality-invariant [14], in the sense that it is the
same for all backgrounds, including those with torsion,
obtained via $O(d,d)$ transformations [6] from the vacuum
dilaton-driven background.

The spectrum (3.1) has also the same
slope (modulo
logarithmic corrections) as the low frequency part of
a thermal black body spectrum, which can be written (in critical
units) as
$$
\Om _T(\om, t)= {\om^4\over \r_c}{1\over e^{\om/ T} -1}
 \simeq B \Om_\ga
\left(H_s\over M_p\right)^2
\left(\om\over \om_s\right)^3 {T\over \om_s} ,  \,\,~~~~~
\om <T\eqno(3.2)
$$
Here $B=(H_s/H_1)^2(a_s/a_1)^4$ is a constant factor which
depends on the time-gap between the beginning of the string
phase and the beginning of the string era. We can thus
parameterize the graviton spectrum (3.1) in terms of an
effective temperature
$$
T_s=(A/B) \om_s \eqno (3.3)
$$
which depends on the initial curvature scale $H_s$, and on the
subsequent kinematic of the high-curvature string phase.

For a negligible duration of the string phase, $t_s\sim t_1$, we
have in particular $H_s \sim H_1 \sim M_p$, and the spectrum
(3.1) is peaked around a maximal amplified frequency $\om_s
\sim H_1 a_1/a \sim 10^{11}$Hz, while it is exponentially
decreasing at higher frequencies (where the parametric
amplification is not effective). Moreover, $T_s \sim \om_s
\sim 1 ^oK$, so that this spectrum, produced by a geometry
transition, is remarkably similar to that of the observed cosmic
black body radiation [17] (see also Sect. 9 below).

The problem, however, is that we don't know the duration and
the kinematics of the high curvature string phase. As a
consequence, we know the slope ($\om ^3$) of this ``dilatonic"
branch of the spectrum, but we don't know the position, in the
($\Om,\om$) plane, of the peak frequency $\om_s$. This
uncertainty is, however, interesting, because the effects of
the string phase could shift the spectrum (3.1) to a low enough
frequency band, so as to overlap with the possible future
sensitivity of
large interferometric detectors such as LIGO [31] and VIRGO
[32]. I will discuss this possibility in terms of a two-parameter
model of background evolution, presented in the following
Section.

\vskip 1.5 cm
\noi
{\bf 4. Two-parameter model of background evolution}
\bigskip
\noi
Consider the scenario described in Sect. 1 (see also [3,8,9]),
in which the initial (flat and cold) vacuum state, possibly
perturbed by the injection of an arbitrarily small (but finite)
density of bulk string matter, starts an accelerated evolution
towards a phase of growing curvature and dilaton coupling,
where the matter contribution becomes eventually negligible
with respect to the gravi-dilaton kinetic energy. Such a phase
is initially described by the low energy dilaton-dominated
solution,
$$
a=|\e|^{1/2} , \,\,\,\,\,\,\,\,\,\,
\phi = -\sqrt 3 \ln |\e| , \,\,\,\,\,\,\,\,\,\,
-\infty < \e < \es \eqno(4.1)
$$
up to the time $\es$, when the curvature reaches the string
scale $H_s=\la_s^{-1}$, at a value of the string coupling $g_s=
\exp (\phi_s/2)$. Provided the value of $\phi_s$ is sufficiently
negative (i.e. provided the coupling $g_s$ is sufficiently
small to be still in the perturbative regime), such a value is
also completely arbitrary, since there is no perturbative
potential to break invariance under shifts of $\phi$.

For $\e >\es$ the background enters a high curvature string
phase of arbitrary (but unknown) duration, in which all
higher-derivative (higher-order in $\ap =\la_s^2$)
contributions to the effective action become important.
During such phase the dilaton keeps growing towards the
strong coupling regime, up to the time $\e=\e_1$ (at a
curvature scale $H_1$), when a non-trivial dilaton potential
freezes the coupling to its present constant value $g_1=
\exp(\phi_1/2)$. We shall assume, throughout this paper, that
the time scale $\e_1$ marks the end of the string era as well
as the (nearly simultaneous) beginning of the standard,
radiation-dominated evolution, where $a\sim \e$ and $\phi =$
const (see however Sect. 9 for a
possible alternative).

During the string phase the curvature is expected to stay
controlled by the string scale, so that
$$
|H|\simeq g M_p = {e^{\phi/2} \over \la_p} = {1\over \la_s} ,
\,\,\,\,\,\,\,~~~~~~ \es < \e < \e_1 \eqno (4.2)
$$
where $\la_p$ is the Planck length. As a consequence, the
curvature is increasing in the Einstein frame (where $\la_p$ is
constant), while it keeps constant in the string frame, where
$\la_s$ is constant and the Planck length grows like $g$ from
zero (at the initial vacuum) to its present value $\la_p\simeq
10^{-19}(GeV)^{-1}$. In both cases the final scale $H_1\simeq
g_1 M_p$ is fixed, and has to be of Planckian order to match
the present value of the ratio $\la_p/\la_s$. Using standard
estimates [33]
$$
g_1 \simeq {H_1\over M_p} \simeq \left({\la_p \over\la_s}\right)_{t_0}
 \simeq
0.3 - 0.03 \eqno(4.3)
$$

In analogy with the dilaton-driven solution (4.1), let us now
parameterize, in the Einstein frame, the background kinematic
during the string phase with a monotonic metric and dilaton
evolution,
$$
a=|\e|^{\a} , \,\,\,\,\,\,\,\,\,\,
\phi = -2\b \ln |\e| , \,\,\,\,\,\,\,\,\,\,
\e_s< \e < \e_1 \eqno(4.4)
$$
representing a sort of ``average" time-behavior. Note that
the two parameters $\a$ and $\b$ cannot be independent
since, according to eq. (4.2),
$$
\left |H_s\over H_1\right| \simeq {g_s\over g_1} \simeq
\left |\e_1\over \es\right|^{1+\a}\simeq
\left |\e_1\over \es\right|^{\b} \eqno(4.5)
$$
from which
$$
1+\a \simeq \b \simeq -{\log (g_s/g_1)\over \log |\es/\e_1|}
\eqno(4.6)
$$
(note also that the condition $1+\a=\b$ cannot be satisfied by
the vacuum
solutions of the lowest order string effective action [3], in
agreement with the fact that all orders in $\ap$ are full
operative in the high curvature string phase [20]).

The background evolution, for this class of models, is thus
completely determined in terms of two parameters only, the
duration (in conformal time) of the string phase, $|\es/\e_1|$,
and the shift of the dilaton coupling (or of the curvature scale
in Planck units) during the string phase, $g_s/g_1=(H_s/M_p)/
(H_1/M_p)$. I will use, for convenience, the decimal logarithm
of these parameters,
$$
\eqalign {x&=\log_{10} |\es/\e_1|=\log_{10} z_s \cr
y&=\log_{10} (g_s/g_1)=\log_{10} {(H_s/M_p)\over (H_1/M_p)}
\cr }
\eqno(4.7)
$$
Here $z_s=|\es/\e_1|\simeq a_1/a_s$ defines the total
red-shift associated to the string phase in the String frame,
where the curvature is approximately constant and the
metric undergoes a
phase of nearly de Sitter expansion. It should be noted, finally,
that the parameters (4.7) are completely frame-independent,
as conformal time and dilaton field are exactly the same both in the
String and Einstein frame.

\vskip 1.5 cm
\noi
{\bf 5. Parameterized graviton spectrum}
\bigskip
\noi
Consider the background discussed in the previous Section,
characterized by the dilaton-driven evolution (4.1) for
$\e<\es$, by the string evolution (4.4) for $\es<\e<\e_1$, and
by the standard radiation-dominated evolution for $\e>\e_1$.
In these three regions, eq. (2.2) for the canonical variable
$u_k$ has the general exact solution
$$
\eqalign{u_k&=|\eta|^{1/2}
H_{0}^{(2)}(|k\eta|), ~~~~~~~~~~~~~~~~~~~~~~~~~~~\,\,\,\,\,\,\,\,\,\,\,\,
{}~~~~~~~~~~~~~\eta<\eta_{s}\cr
u_{k} &= |\eta|^{1/2}\left[ A_{+}(k)
H_{\nu}^{(2)}(|k\eta|) + A_{-}(k)
H_{\nu}^{(1)} (|k\eta|)\right] ,~~~~~~~~~
\eta_{s}<\eta<\eta_{1} \cr
u_{k}&= {1\over \sqrt k}\left[c_{+}(k)e^{-ik\eta}
+c_{-}(k)e^{ik\eta}\right] ,~~~~~~~~~~~~~~~~~~~~~~~~~~~\eta>\eta_{1} \cr}
\eqno(5.1)
$$
where $\nu=|\a-1/2|$, and
$H^{(1,2)}$
are the first and second kind Hankel functions. We have
normalized the solution to an initial vacuum fluctuation
spectrum, containing only positive frequency modes at $\e
=-\infty$
$$
\lim_{\e \ra -\infty} u_k= {e^{-ik\e}\over \sqrt k} \eqno(5.2)
$$
The asymptotic solution for $\e \ra +\infty$ is however a linear
superposition of positive and negative frequency modes,
determined by the so-called Bogoliubov coefficients
$c_\pm(k)$ which parameterize, in a second quantization
approach, the unitary transformation connecting $|in\rangle $
and $|out \rangle$ states. So, even starting from an initial
vacuum state, it is possible to find a non-vanishing expectation
number of produced particles (in this case gravitons) in the
final state, given (for each mode $k$) by $\langle n_k \rangle=
|c_-(k)|^2$.

We shall compute $c_\pm$ by matching the solutions (5.1) and
their first derivatives at $\es$ and $\e_1$. We observe, first of
all, that the required growth of the curvature and of the coupling
during the string phase (in the Einstein frame) can only be
implemented for $|\e_1|<|\es|$, i.e. $\b=1+\a>0$ [see
eq.(4.5)]. This leads to an inflationary string phase,
characterized in the Einstein frame by accelerated expansion
($\dot a>0$, $\ddot a >0$, $\dot H >0$) for $-1<\a<0$, and
accelerated contraction
($\dot a<0$, $\ddot a <0$, $\dot H <0$) for $\a>0$.
As a consequence, modes which ``hit"
the effective potential barrier $V(\e)=a''/a$ of eq.(2.2)
(otherwise stated: which cross the horizon) during the
dilaton-driven phase, i. e. modes with $|k\e_s|<1$, stay under
the barrier also during the string phase, since $|k\e_1|< |k\e_s|
<1$. In such case the maximal amplified proper frequency
$$
\om_1 ={k_1\over a}\simeq {1\over a\e_1}
\simeq {H_1a_1\over a}\simeq \left(H_1\over M_p\right)^{1/2}
10^{11} Hz
=\sqrt {g_1} 10^{11} Hz \eqno(5.3)
$$
is related to the highest frequency crossing the horizon in the
dilaton phase, $\om_s=H_s a_s/a$, by
$$
\om_s=\om_1 \left|\e_1\over \es\right| <\om_1 \eqno(5.4)
$$
For an approximate estimate of $c_-$ we may thus consider two
cases.

If $\om_s<\om < \om_1$, i.e. if we consider modes crossing the
horizon in the string phase, we can estimate $c_-(\om)$ by
using the large argument limit of the Hankel functions when
matching the solutions at $\e=\es$, using however the small
argument limit when matching at $\e=\e_1$. In this case the
parametric amplification is induced by the second background
transition only, as $A_+\simeq 1$ and $A_-\simeq 0$, and we
get
$$
|c_-(\om)|\simeq \left(\om\over
\om_1\right)^{-\nu-1/2} , \,\,\,\,\,\,\,\,\,\,~~~~~
\om_s<\om<\om_1
\eqno(5.5)
$$
(modulo numerical coefficients of order of unity). If, on the
contrary, $\om<\om_s$, i.e. we consider modes crossing the
horizon in the dilaton phase, we can use the small argument
limit of the Hankel functions at both the matching epochs $\es$
and $\e_1$. This gives $A_\pm=b_\pm |k\es|^{-\nu}\ln |k\es|$
($b_\pm$ are numbers of order one), and
$$
|c_-(\om)|\simeq \left |\es\over \e_1\right|^\nu
\left(\om\over
\om_1\right)^{-1/2}\ln \left(\om_s\over\om\right) , \,\,\,\,\,
{}~~~~~\om<\om_s
\eqno(5.6)
$$

We can now compute, in terms of $\langle n \rangle=
|c_-|^2$, the spectral energy distribution $\Om(\om,t)$ (in
critical units) of the produced gravitons, defined in such a way
that the total graviton energy density $\r_g$ is obtained as
$\r_g=\r_c\int \Om(\om)d\om/\om$. We have then
$$
\eqalign{\Om(\om,t)&\simeq {\om^4\over
M_p^2H^2}|c_-(\om)|^2 \simeq \cr
&\simeq \left(H_1\over M_p\right)^2
\left(H_1\over H\right)^2\left(a_1\over a\right)^4
\left(\om\over \om_1\right)^{3-2\nu} , ~~~~~~
\,\,\,\,\,\,~~~ \om_s<\om<\om_1 \cr
&\simeq \left(H_1\over M_p\right)^2
\left(H_1\over H\right)^2\left(a_1\over a\right)^4
\left(\om\over \om_1\right)^{3} \left|\es\over
\e_1\right|^{2\nu}\ln^2\left(\om_s\over \om \right) ,~~~
\,\, \om< \om_s \cr}\eqno(5.7)
$$
According to eqs. (4.6) and (4.7), moreover, $2\nu=|2\a -1|=
|3+2y/x|$ and $|\es/\e_1|=10^x$. The tensor perturbation
spectrum (5.7) is thus completely fixed in terms of our two
free parameters $x, y$, of the (known) fraction of critical
energy density stored in radiation at time $t$, $\Om_\ga (t)=
(H_1/H)^2(a_1/a)^4$, and of the (in principle known) present
value of the ratio $g_1=\la_p/\la_s$, as
$$
\Om(\om,t)=g_1^2\Om_\ga (t)\left(\om\over
\om_1\right)^{3-|{2y\over x}+3|} , ~~~~\,\,\,\,\,\, 10^{-x}<{\om
\over \om_1}<1  \eqno(5.8)
$$
$$
\Om(\om,t)=g_1^2\Om_\ga (t)\left(\om\over
\om_1\right)^{3}10^{|{2y}+3x|} , ~~~~\,\,\,\, \,\,{\om
\over \om_1}<10^{-x} \eqno(5.9)
$$
The same spectrum can also be
obtained, with a different approach, working directly in the String
frame (see [14]).

The first branch
of this spectrum, with unknown slope, is due to modes crossing
the horizon in the string phase, the second to modes crossing
the horizon in the dilaton phase (I have omitted, for
simplicity, the logarithmic term in eq. (5.9), because it is not
much relevant for the order of magnitude estimate that I want
to discuss here). Note that, as previously stressed, the cubic slope of
the dilatonic branch of the spectrum is completely insensitive to the
kinematic details of the string phase. Such details can only affect the
overall intensity of the perturbation spectrum, and their effects can
thus be absorbed by rescaling the total duration of the string phase.
For the case $\a<1/2$, in which the amplitude of perturbations is
constant outside the horizon, we recover in particular eq.(3.1) with $A=
1$.

Let us now impose on such spectrum the condition of falling within
the possible future sensitivity range of large interferometric
detectors (such as that of the ``Advanced LIGO" project [34]),
namely
$$
\Om(\om_I) \gaq 10^{-10} , \,\,\,\,\,\,\,\,\,\,\,
\om_I =10^2 Hz \eqno(5.10)
$$
It implies
$$
|y+{3\over 2}x|> {3\over 2}x -{(3+\log_{10}g_1)x\over 9+
\log_{10}g_1} , \,\,\,\,\,\, x> 9+{1\over2}\log_{10}g_1
$$
$$
|2y+3x|>21-{1\over 2}\log_{10}g_1 , \,\,\,\,\,\,\,\,\,\,
x<9+{1\over 2}\log_{10}g_1 \eqno(5.11)
$$
These conditions define an allowed region in our
parameter space $(x,y)$ which has to be further restricted,
however, by the upper bound obtained from pulsar-timing
measurements [35], namely
$$
\Om(\om_P) \laq 10^{-6} , \,\,\,\,\,\,\,\,\,\,\,
\om_P =10^{-8} Hz \eqno(5.12)
$$
which implies
$$
|y+{3\over 2}x|< {3\over 2}x -{(1+\log_{10}g_1)x\over
19+{1\over 2} \log_{10}g_1} , \,\,\,\,\,\, x>
19+{1\over2}\log_{10}g_1
$$
$$
|2y+3x|<55-{1\over 2}\log_{10}g_1 , \,\,\,\,\,\,\,\,\,\,
x<19+{1\over 2}\log_{10}g_1 \eqno(5.13)
$$

We have to take into account, in addition, the asymptotic
behavior of tensor perturbations outside the horizon. During
the dilaton phase the growth is only logarithmic, but during
the string phase the growth is faster (power-like) for
backgrounds with $\a>1/2$. Since the above spectrum has been
obtained in the linear approximation, expanding around a
homogeneous background, we must impose for consistency
that the perturbation amplitude stays always smaller than
one, so that perturbations have a negligible back-reaction on
the metric. This implies $\Om<1$ at all $\om$ and $t$. This
bound, together with the slightly more stringent bound
$\Om<0.1$ required by standard nucleosynthesis [36], can be
automatically satisfied - in view of the $g_1^2$ factor in eqs.
(5.8), (5.9) - by requiring a growing perturbation spectrum,
namely
$$
y<0 , \,\,\,\,\,\,\,\,\,\,\,\,\,\,\,\,~~~~~~~~~~~
 y>-3x \eqno(5.14)
$$

The conditions (5.11), (5.13) and (5.14) determine the allowed
region of our parameter space compatible with the production
of cosmic gravitons in the interferometric sensitivity range (5.10)
(denoted by LIGO, for short). Such a region is plotted in {\bf
Fig.1}, by taking $g_1=1$ as a reference value. It is bounded
below by the condition of nearly homogeneous
background (5.14), and above by the same condition plus the
pulsar bound (5.13). The upper part of the allowed region
corresponds to a class of backgrounds in which the tensor
perturbation amplitude stays constant, outside the horizon,
during the string phase ($\a<1/2$). The lower part corresponds
instead to backgrounds in the the amplitude grows, outside the
horizon, during the string phase ($\a>1/2$).
The area within the full bold lines refers
to modes crossing the horizon in the dilaton phase; the area
within the thin lines  refers
to modes crossing the horizon in the string phase, where the
reliability of our predictions is weaker, as we used
field-theoretic methods in a string-theoretic regime. Even
neglecting all spectra referring to the string phase, however,
we obtain a final allowed region which is non-vanishing,
though certainly not too large.

The main message of this figure and of the spectrum
(5.7) (irrespective of the particular value of the spectral
index) is that graviton production, in string cosmology, is in
general strongly enhanced in the high frequency sector
(kHz-GHz). Such a frequency band, in our context, could be in
fact all but the ``desert" of relic gravitational radiation that
one may expect on the ground of the standard inflationary
scenario. This conclusion is independent of the kinematical details of
the string phase, which can only affect the shape of the high frequency
tail of the spectrum, but not the peak intensity of the spectrum,
determined by the fundamental string parameter $\la_s$. As a
consequence,
a sensitivity of $\Om \sim
10^{-4} - 10^{-5}$ in the KHz region (which does not seem out of reach
in coincidence experiments between bars and interferometers [37])
could be already
enough to detect a signal, so that a
null result (in that
band, at that level of sensitivity) would already provide a
significant constraint on the parameters of the string
cosmology background. This should encourage the study and
the development of gravitational detectors (such as, for
instance, microwave cavities [38]) with large sensitivity in the
high frequency sector.

In the following Section I will compare the allowed region of
{\bf Fig. 1}, relative to graviton production (and their possible
detection), to the allowed region relative to the amplification
of electromagnetic perturbations (and to the production of
primordial magnetic fields).

\vskip 1.5 cm
\noi
{\bf 6. Parameterized electromagnetic spectrum}
\bigskip
\noi
In string cosmology, the electromagnetic field $F_{\mu\nu}$ is
directly coupled to the dilaton background. To lowest order,
such coupling is represented by the string effective action as
$$
\int d^{d+1}x\sqrt{|g|} e^{-\phi}F_{\mu\nu}F^{\mu\nu}
\eqno(6.1)
$$
The electromagnetic field is also coupled to the metric
background $g_{\mu\nu}$, of course, but in $d=3$ the
metric coupling is conformally invariant, so that no parametric
amplification of the electromagnetic fluctuations is possible in a
conformally flat background, like that of a typical inflationary
model. One can try to break conformal invariance at the
classical or quantum level - there are indeed various attempts
in this sense [39,40] - but it turns out that it is very difficult, in
general, to obtain a significant electromagnetic amplification
from the metric coupling in a natural way, and in a realistic
inflationary scenario.

In our context, on the contrary, the vacuum fluctuations of the
electromagnetic field can be directly amplified by the time
evolution of the dilaton background [15,41]. Consider in fact the
correct canonical variable $\psi^\mu$ representing
electromagnetic perturbations [according to eq. (6.1)] in a
$d=3$, conformally flat background, i.e. $\psi^\mu=A^\mu
e^{-\phi/2}$, where $F_{\mu\nu}= \pa_\mu A_\nu -
\pa_\nu A_\mu$. The Fourier modes $\psi^\mu_k$ satisfy, for
each polarization component, the equation
$$
\psi_k''+\left[k^2-V(\e)\right]\psi_k=0 , \,\,\,\,\,\,\,\,
{}~~~~~~V(\e)= e^{\phi/2}(e^{-\phi/2})'' \eqno(6.2)
$$
obtained from the action (6.1) by imposing the standard radiation
gauge for electromagnetic waves in vacuum, $A^0=0$,
${\bf \nabla \cdot A}=0$.
Such equation is very
similar to the tensor perturbation equation (2.2), with the only
difference that the Einstein scale factor, in the effective potential
$V(\eta)$, is replaced by the
inverse of the string coupling $g^{-1}=e^{-\phi/2}$.

Consider now the string cosmology background of Sect. 4, in
which the dilaton-driven phase (4.1) and the string phase (4.4)
are followed by the radiation-driven expansion. For such
background, the effective potential (6.2) is given explicitly by
$$
\eqalign{
V&={1\over {4\eta^2}} (3 -
\sqrt{12}) ,~~~~~~~~~~~~~~~\,\,\,\,\eta<\eta_{s}  \cr
V&={\beta(\beta-1)\over{\eta^2}} ,
{}~~~~~~~~~~~~~~~~\eta_{s}<\eta<\eta_{1} \cr
V&=0 ,~~~~~~~~~~~~~~~~~~~~~~~~~~~~~~~~~\eta>\eta_{1}\cr}
\eqno(6.3)
$$
The exact solution of eq. (6.2), normalized to an initial
vacuum fluctuation spectrum ($\psi_k \ra e^{-ik\e}/\sqrt k$ for
$\e \ra -\infty$), is thus
$$
\eqalign{\psi_k&=|\eta|^{1/2}
H_{\sg}^{(2)}(|k\eta|) ,~~~~~~~~~~~~~~~~~~~~\,\,\,\,\,\,\,\,\,\,\,\,
{}~~~~~~~~~~~~~~~~\eta<\eta_{s}\cr
\psi_{k} &= |\eta|^{1/2}\left[ B_{+}(k)
H_{\mu}^{(2)}(|k\eta|) + B_{-}(k)
H_{\mu}^{(1)} (|k\eta|)\right] ,~~~~~
\eta_{s}<\eta<\eta_{1} \cr
\psi_{k}&= {1\over \sqrt k}\left[c_{+}(k)e^{-ik\eta}
+c_{-}(k)e^{ik\eta}\right] ,~~~~~~~~~~~~~~~~~~~~~~~\eta>\eta_{1} \cr}
\eqno(6.4)
$$
where $\sg= (\sqrt 3 -1)/2$, and $\mu=|\b-1/2|$.

For this model of background the effective potential
grows in the dilaton phase, keeps growing in the
string phase where it reaches a maximum $\sim \e_1^{-2}$
around the transition scale $\e_1$, and then goes rapidly to
zero in the subsequent radiation phase, where
$\phi=\phi_1=$const. The maximum amplified frequency is of
the same order as before,
$\om_1=H_1a_1/a=|\es/\e_1|\om_s >\om_s$, where
$\om_s=H_sa_s/a$ is the last mode hitting the barrier (or
crossing the horizon) in the dilaton phase. For modes with
$\om>\om_s$ the amplification is thus due to the second
background transition only: we can evaluate $|c_-|$ by using
the large argument limit of the Hankel functions when
matching the solutions at $\es$ (which gives $B_+\simeq 1$,
$B_-\simeq 0$), using however the small argument limit when
matching at $\e_1$, which gives
$$
|c_-(\om)|\simeq \left(\om\over \om_1\right)^{-\mu-1/2} ,
\,\,\,\,\,\,\,\, \om_s<\om<\om_1 \eqno (6.5)
$$
Modes with $\om<\om_s$, which exit the horizon in the dilaton
phase, stay outside the horizon also in the string phase, so
that we can use the small argument limit at both the matching
epochs: this gives $B_\pm =b_\pm |k\es|^{-\sg -\mu}$
($b_\pm$ are numbers of order of unity) and
$$
|c_-(\om)|\simeq \left(\om \over \om_s \right)^{-\sg}
\left(\om\over \om_1\right)^{-1/2}\left|\es\over
\e_1\right|^{\mu} ,  \,\,\,\,\,\, \om<\om_s \eqno (6.6)
$$

We are interested, in particular, in the ratio
$$
r(\om)={\om\over \r_{\ga}}{d\r\over d\om}\simeq
{\om^4\over \r_{\ga}}|c_-(\om)|^2 \eqno(6.7)
$$
measuring the fraction of electromagnetic energy density
stored in the mode $\om$, relative to the total radiation
energy $\r_\ga$. By using the parameterization of Sect. 4 we
have $2\mu=|2\b-1|=|1+2y/x|$ and $|\es
/\e_1|=\om_1/\om_s=10^x$, so that the electromagnetic
perturbation spectrum is again determined by two parameters
only, the duration of the string phase $|\es/\e_1|$, and the
initial value of the string coupling, $g_s=g_1 10^y$. We find
$$
r(\om)=g_1^2\left(\om\over
\om_1\right)^{3-|{2y\over x}+1|} , ~~~~~
\,\,\,\,\,\, 10^{-x}<{\om
\over \om_1}<1  \eqno(6.8)
$$
for modes crossing the horizon in the string phase, and
$$
r(\om)=g_1^2\left(\om\over
\om_1\right)^{4-\sqrt 3}10^{x(1-\sqrt 3) +|{2y}+x|} ,
\,\,\,\,~~~~~
\,\,{\om \over \om_1}<10^{-x} \eqno(6.9)
$$
for modes crossing the horizon in the dilaton phase. The same
spectrum has been obtained, with a different approach, also in
the String frame [15,16] (in this paper, the definition of
$g_1$ has been rescaled with respect to [16],
by absorbing into $g_1$ the
$4\pi$ factor). Finally, this spectrum can be easily generalized to
include the effects of an arbitrary number of shrinking internal
dimensions during the pre-big-bang phase [15]. The basic qualitative
result presented in the following Sections hold however independently of
such a generalization.
\vskip 1.5 cm
\noi
{\bf 7. Seed magnetic fields}
\bigskip
\noi
The above spectrum of amplified electromagnetic vacuum
fluctuations has been obtained in the linear approximation,
expanding around a homogeneous background. We have thus to
impose on the spectrum the consistency condition of negligible
back-reaction, $r(\om)<1$ at all $\om$.  For $g_1<1$ this
condition requires a growing perturbation spectrum, and
imposes a rather stringent bound on parameter space,
$$
y<x , \,\,\,\,\,\,\,\,\,\,\,\,\,\,\,\,\, y>-2x \eqno(7.1)
$$
[note that a growing spectrum also automatically satisfies the
nucleosynthesis bound $r<0.1$, in view of the $g_1^2$ factor
which normalizes the strength of the spectrum, and of eq.
(4.3)].

It becomes now an interesting question to ask whether, in
spite of the above condition, the amplified
fluctuations can be large enough to seed the dynamo
mechanism which is widely believed to be responsible for the
observed galactic (and extragalactic) magnetic fields [42]. Such
a mechanism would require a primordial magnetic field
coherent over the intergalactic Mpc scale, and with a
minimal strength such that [39]
$$
r(\om_G) \gaq 10^{-34} , \,\,\,\,\,\,\,\,\,\,\,\, \om_G =10^{-14}
Hz \eqno(7.2)
$$
This means, in terms of our parameters,
$$
|y+{x\over2}|>{3\over2}x-{(17+\log_{10}g_1)x \over 25+
{1\over2}\log_{10}g_1} , \,\,\,\,\, x>25+{1\over2}\log_{10}g_1
$$
$$
x(1-\sqrt 3)+|2y+x|>23-0.87 \log_{10}g_1 , \,\,\,\,\,
x<25+{1\over2}\log_{10}g_1 \eqno(7.3)
$$
Surprisingly enough the answer to the previous question is
positive, and this marks an important point in favor of the
string cosmology scenario considered here, as it is in general
quite difficult - if not impossible -  to satisfy the condition
(7.2) in other, more conventional, inflationary scenarios.

The allowed region of parameter space, compatible with the
production of seed fields [eq. (7.3)] in a nearly homogeneous
background [eq. (7.1)], is shown in {\bf Fig. 2} (again for the
reference value $g_1=1$). In the region within the full bold lines
the seed fields are due to modes crossing the horizon in the
dilaton phase, in the region within the thin lines to modes
crossing the horizon in the string phase. In both cases the
background satisfies $y<-x/2$, i.e. $\b>1/2$, so that the whole
allowed region refers to perturbations which are always
growing outside the horizon, even in the string phase.

We may see from {\bf Fig. 2} that the production of seed fields
require a very small value of the dilaton coupling at the
beginning of the string phase,
$$
g_s= e^{\phi_s/2} \laq 10^{-20}\eqno(7.4)
$$
This initial condition is particularly interesting, as it could have
an important impact on the problem of freezing out the
classical oscillations of the dilaton background (work is in
progress). It also requires a long enough duration of the string
phase,
$$
z_s=|\es/\e_1|\gaq 10^{10} \eqno(7.5)
$$
\noi
which is not unreasonable, however, when $z_s$ is translated
in cosmic time and string units, $z_s=\exp(\Da t/\la_s)$, namely
$\Da t \gaq 23\la_s$.

Also plotted in {\bf Fig. 2}, for comparison, are the allowed
regions for the production of gravitons falling within the
interferometric sensitivity range, taken from the previous
picture. Since there is no overlapping, a signal detected (for
instance) by LIGO would seem to exclude the possibility of
producing seed fields, and vice-versa. Such a conclusion should
not be taken too seriously, however, because the allowed
regions of {\bf Fig. 2} actually define a ``minimal" allowed area,
obtained within the restricted range of parameters compatible
with a linearized description of perturbations. If we drop the
linear approximation, then the allowed area extends to the
``south-western" part of the plane ($x,y$), and an overlap
between electromagnetic and gravitational regions becomes
possible. In that case, however, the perturbative approach
around a homogeneous background could not be valid any
longer, and we would not be able to provide a correct
computation of the spectrum.

The above discussion is based on the low-energy string effective action
(6.1). Of course there are corrections to this action coming from the
loop expansion in the string coupling $g=e^{\phi/2}$,
and higher curvature
corrections coming from the $\ap$ expansion. Loop corrections are not
important, however, until we work in a region of parameter space in
which the dilaton is deeply in its perturbative regime ($g<<1$), as is
indeed the case for the production of seed fields (moreover,
the non-perturbative dilaton potential is known to be extremely small,
as $V(\phi)\sim \exp(-1/g^2)$ in the weak coupling regime [9]). The
$\ap$ corrections may be important, but only for modes which crossed the
horizon during the string phase (for modes crossing the horizon
before the slope of the spectrum is unaffected by the subsequent
background kinematics, as already stressed in Sect. 5 for the case of
tensor perturbations). It should be clearly stressed, therefore, that
the main result of this Section - namely the existence of a wide
region of parameter space in which seed fields are efficiently
produced - is completely independent of the unknown details of the
string phase, which can influence in a direct way only the high
frequency tail of the spectrum, and only control the possible extension
of the allowed region in the limit of very large $z_s$.

\vskip 1.5 cm
\noi
{\bf 8. The anisotropy of the CMB radiation}
\bigskip
\noi
In the inflationary models based on the low energy string
effective action, the spectrum of scalar and tensor metric
perturbations grows in general too fast with frequency
[3,12,13] to be able to explain the large scale anisotropy
detected by COBE [43,44]. If we insist, however, in looking for
an explanation of the anisotropy in terms of the quantum
fluctuations of some primordial field (amplified by the
background evolution), a possible - even if unconventional -
explanation in a string cosmology context is provided by the
vacuum fluctuations of the electromagnetic field [16].

Consider in fact the electromagnetic perturbations re-entering
the horizon ($|k\e|\sim 1\sim \om/H$) after amplification. At
the time of reentry $H^{-1}$ they provide a field coherent over
the horizon scale, which can seed the cosmic magnetic fields, as
discussed in the previous Section. If reentry occurs before the
decoupling era, the perturbations may be expected to thermalize and to
become homogeneous very rapidly soon after reentry,
because of
their interactions with matter sources in thermal equilibrium.
Modes crossing the horizon after decoupling, on the
contrary,  contribute to the
formation of a stochastic perturbation background
with a spectrum which remains frozen until the present time $t_0$,
and which may significantly affect the isotropy and homogeneity of
the CMB radiation.
For a complete electromagnetic
origin of the observed
anisotropy, $\Da T/T$, at the present horizon scale,
$\om_0\sim 10^{-18}$Hz, the perturbation amplitude should
satisfy in particular the condition
$$
r(\om_0)\Om_\ga (t_0) \sim (\Da T/T)^2_0
\eqno(8.1)
$$
namely
$$
r(\om_0)\simeq 10^{-6} , \,\,\,\,\,\,\,\,\,\, \om_0= 10^{-18} Hz
\eqno(8.2)
$$

According to our spectrum [eqs.(6.8) and (6.9)]
this condition can be satisfied consistently with the
homogeneity bound (7.1), and without fine-tuning of
parameters, provided the string phase is so long that all scales
inside our present horizon crossed the horizon (for the first
time) during the string phase, i. e. for $\om_0>\om_s$ (or
$z_s>10^{29}$). If we accept this electromagnetic explanation
of the anisotropy, we have then two important consequences.

The first follows from the fact that the peak value of the
spectrum (6.8) is fixed, so that the spectral index $n$, defined
by
$$
r(\om)= g_1^2 \left(\om\over \om_1\right)^{n-1} \eqno (8.3)
$$
can be completely determined as a function of the amplitude
at a given scale. For the horizon scale, in particular, we have
from eqs. (8.1) and (5.3)
$$
n\simeq {25+{5\over 2}\log_{10}g_1- 2 \log_{10}(\Da
T/T)_{\om_0} \over 29 +{1\over 2} \log_{10}g_1} \eqno (8.4)
$$
I have taken explicitly into account here the dependence of
the spectrum on the the present value of the string coupling
$g_1$ (which is illustrated in {\bf Fig. 3}), to stress that such
dependence is very weak, and that our estimate of $n$ from
$\Da T/T$ is quite stable, in spite of the rather large theoretical
uncertainty about $g_1^2$ (nearly two order of magnitude,
recall eq. (4.3)).

In order to match the observed anisotropy, $\Da T/T \sim
10^{-5}$, we obtain from eq. (8.4) (see also {\bf Fig. 3}, where the
relation (8.4) is plotted for three different values of $g_1$)
$$
n \simeq 1.11 - 1.17 \eqno(8.5)
$$
This slightly growing (also called ``blue" spectrum)
is flat enough to be well compatible with the present analyses
of the COBE data [43,44].

The second consequence follows from the fact that fixing a
value of $n$ in eq. (8.3) amounts to fix a relation between the
parameters $x$ and $y$ of our background, according to eq.
(6.8). If we accept, in particular, a value of $n$ in the range of
eq. (8.5), then we are in a region of parameter space which is
also compatible with the production of seed fields, according
to eq. (7.2). This means that we are allowed to formulate
cosmological models in which cosmic magnetic fields and CMB
anisotropy have the same common origin, thus
explaining (for instance) why the energy density $\r_B$ of the
observed cosmic magnetic fields is of the same order as that
of the CMB radiation:
$$
\r_B \sim \r_\ga \int^{\om_1} r(\om)d(\ln \om) \sim \r_{CMB}
\eqno(8.6)
$$
A coincidence which is otherwise mysterious, to the best of my
knowledge.

It is important to mention that the values of the parameters
leading to eq.(8.5) are also automatically consistent with the
bound following from the presence of strong magnetic fields at
nucleosynthesis time [45], which imposes $r(\om_N)\laq 0.05$
at the scale corresponding to the end of nucleosynthesis,
$\om_N \simeq 10^{-12}$Hz. By comparing photon and graviton
production [eqs. (6.8) and (5.8)] we find, moreover, that for a
background in which $n$ lies in the range (8.5) the graviton
spectrum grows
fast enough with frequency ($\Om \sim
\om^{m}$, $m=n+1= 2.11 - 2.17$) to be well compatible with
the pulsar bound (5.12). Note that, with such a value of $m$,
the metric perturbation contribution to the COBE anisotropy is
completely negligible.

It should be stressed, finally, that (in contrast with what discussed in
the previous Section) the details of the string phase are of crucial
importance for a possible electromagnetic explanation of the CMB
anisotropy. However, once we accept a model in which the background
curvature stays nearly constant in the String frame, the anisotropy
discussed in this Section is again generated in a range of parameters
for which the dilaton is deeply inside the perturbative regime ($g<<1$),
so that the electromagnetic perturbation equations are certainly stable
with respect to loop corrections. Moreover, since we are expanding
around the vacuum background ($F_{\mu\nu}=0$), the perturbation
equations are also stable in the linear approximation against $\ap$
corrections, provided such corrections appear in the form of powers of
the Maxwell field strength with no higher derivative term. This
behaviour, on the other hand, is typical of the Eulero forms (like the
Gauss-Bonnet invariant) which are conjectured to contribute to the
correct (ghost-free) higher curvature expansion of the gravity sector
[46]. The possibility discussed in this Section can thus find consistent
motivations in a string cosmology context.

\vskip 1.5 cm
\noi
{\bf 9. Possible origin of the CMB radiation}
\bigskip
\noi
In the standard inflationary picture the
density perturbations of the homogeneous and isotropic
cosmological model,  and the $3^o K$ thermal radiation
background, have physically distinct origin. The former arise from
the vacuum fluctuations of the metric (and possibly other fields),
amplified by the external action of the cosmological background,
which performs a transition from an inflationary phase to a
phase of decelerated evolution. The latter arises instead from the
reheating era subsequent to inflation, with a production
mechanism which is strongly model-dependent [47]  (for instance,
collisions of bubbles produced in the phase transition, and/or
inflaton decay). As first pointed out by Parker [48], however, also
the thermal black-body radiation could have a geometric origin,
with a production mechanism closely related to that
which amplifies inhomogeneities. Indeed, in a string cosmology context,
the class of backgrounds able to provide
an electromagnetic explanation of the CMB anisotropy can
also account for the production of the CMB radiation itself,
directly from the amplification of the vacuum fluctuations of
the electromagnetic (and other gauge) fields.

Without introducing ``ad hoc" some radiation source,
suppose in fact that the background accelerates
up to some maximum (nearly Planckian) scale $H_1$,
corresponding to the peak of the effective potential $V(\e)$ in
the perturbation equations, and then decelerates, with
corresponding decreasing of the potential barrier.
In such a context, the modes of comoving frequency $k$ which ``hit" the
effective potential barrier $|V(\eta)|$
(namely with $|k\eta_1|\laq1$), are parametrically
amplified  by  the external ``pump" field. In a second quantization
language this corresponds, as discussed in Sect. 5,
to a copious production of particles
with a power-like spectral distribution [10],
$$
\langle n_k\rangle=|c_-(k)|^2\simeq |k\eta_1|^{-\a}\eqno(9.1)
$$
(the power $\a$ depends on the slope of the potential barrier,
and then on the kinematical behavior of the background).
The production of particles is instead
exponentially suppressed for those modes which never hit the
barrier,  $|k\eta_1|\gaq 1$. In that case one obtains [48,49]
$$
{|c_-(k)|^2\over 1+ |c_-(k)|^2}\simeq e^{-k\eta_T}\eqno(9.2)
$$
where $\eta_T$ is the scale of (conformal) time characterizing the
transition from the accelerated to the decelerated regime.
For a transition occurring at the time $\eta_1$, in particular, it is
natural to have $\eta_T\sim |\eta_1|$.
For $|k\eta_1|\gaq 1$, all the produced particles are thus
characterized by a distribution of thermal type,
$$
\langle n_k\rangle=|c_-(k)|^2\simeq {1\over
e^{ k|\eta_1|}-1}\eqno(9.3)
$$
as first noted by Parker [48], at a proper temperature $T_1(t)$
determined by the transition curvature scale, $H_1$, as
$$
T_1(t)\sim {1\over  a |\eta_1|}\sim
{H_1a_1\over  a(t)}\eqno(9.4)
$$

The change in the
background evolution thus leads to the production of a mixture of all
kinds of  ultra-relativistic particles, with a spectrum which is
thermal (at a  temperature $T_1$) at high frequency ($\om>T_1$), and
possibly distorted by  parametric amplification effects at low
frequency. For a typical (smaller than Planckian) inflationary scale,
the low frequency part of the spectrum remains frozen
for those particles (like gravitons and dilatons) which interact only
gravitationally, and then decouple immediately after the
background transition; on the contrary, the spectrum at low
frequency may be expected to thermalize rapidly for all the
other produced particles which go on interacting among
themselves (and with the background sources) for a long enough
period of time after the transition. For such particles the spectrum
eventually approaches a thermal distribution in
the whole amplified frequency
band, with a total their energy density $\Om_T$ (in critical units)
fixed by $T_1$ as
$$
\Om_T(t)\sim {GT_1^4\over H^2}\sim
\left(H_1\over M_p\right)^{2}\left(H_1\over
H\right)^{2}\left(a_1\over a\right)^4\eqno(9.5)
$$
Even if, initially, $\Om_T<1$ (as $H_1<M_p$),
this thermal component of the
produced radiation may then become dominant provided,
at the beginning of the decelerated epoch, the scale factor $a(t)$
grows in time more slowly than $H^{-1/2}$ (this is the case, for
instance, of the time-reversed dilaton-dominated solution (4.1) which
expands like $a\sim t^{1/3}$ for $t \ra + \infty$).
In that case the
relics of such radiation might be identified with the (presently
observed) cosmic thermal background, with a
red-shifted temperature $T_1(t_0)\simeq 3^0 K$.

It is important to stress that, if the thermal radiation produced in
the transition becomes a dominant source of the background, such
identification is always possible, in principle, quite independently of
the kinematic details of the inflationary phase, of the transition
scale $H_1$, and of the scale $H_r=H_1(H_1/M_p)(a_1/a_r)^2$ at
which the radiation becomes dominant. In this context, however,
the scale $H_1$ also determines the amplitude of those
fluctuations whose spectrum is not
thermalized at low frequencies, but remains frozen after the
transition. Therefore, if we identify the observed thermal radiation
with that produced in the transition, then the energy density at the
scale $\om_T=T_1$ turns out to be uniquely fixed, also for the
decoupled perturbations, in terms of the energy density of the
observed CMB spectrum. By calling $\Om_P(\om,t)\simeq
\om^4\langle n(\om)\rangle /M_p^2H^2$ the energy density (in
critical units) of such perturbations, we have in fact from eqs. (9.1),
(9.4) and (9.5)
$$
\Om_P(\om_T,t)\sim
\Om_T(t)\eqno(9.6) $$

At the present time $t=t_0$ we thus obtain the constraint
$\Om_P(\om_T)\sim 10^{-4}$ at $\om_T\sim 10^{11}$Hz (modulo
factors of order of unity). This condition must be satisfied, in
particular, by the energy density stored in gravitational (tensor)
perturbations which, as discussed in Sect. 5, is constrained to be
much smaller, $\Om<<10^{-4}$, at lower frequencies (the large scale
degree of isotropy [43,44] implies, for instance,
$\Om_P\laq 10^{-10}$ at $\om\simeq
10^{-18}$Hz). The identification of the observed thermal radiation
with that produced in an inflationary background transition is thus
compatible with such phenomenological bounds, only if the
transition amplifies perturbations with a growing spectrum (which is
indeed the case for the string cosmology scenario discussed here).

The discussion of an explicit example in which thermal radiation is
produced, and eventually becomes dominant, would require however
a model of
smooth background evolution. Such a model cannot be constructed to
the lowest order in $\ap$ from the effective gravi-dilaton action, even
including an arbitrary dilaton potential [20]. It may be constructed,
however,
by including the antisymmetric torsion tensor (equivalent to
the axion in four dimensions)  among the string
background fields, at least if we accept a model of background which is
initially contracting at the beginning of the decelerated regime.
There is no compelling reason, after all, why the phase
of decelerated  expansion should start
immediately after the change from the negative to the positive time
branch of a solution of the string cosmology equations. In particular, a
background transition from accelerated expansion to
decelerated contraction is also an
efficient source of radiation, whose energy density is
naturally led to become dominant,
as shown in the following example
(in an
expansion $\ra$ expansion transition it is instead more
difficult, for the produced
radiation, to dominate over other conventional
background sources).

Consider in fact the background field equations of
motion [50], at tree-level in the string loop expansion parameter
$e^\phi$, and to zeroth order in
$\ap$, written in the String frame
$$
R_\mu^\nu +\bigtriangledown_\mu\bigtriangledown^\nu
\phi- {1\over 2}{\pa V \over \pa \phi}  \da_\mu^\nu  -{1\over
4} H_{\mu\a \b}H^{\nu\a\b} = 8\pi G e^\phi T_\mu^\nu \
$$
$$
R-(\bigtriangledown_\mu\phi)^2+2 \bigtriangledown_\mu
\bigtriangledown^\mu \phi +V-{\pa V\over \pa \phi}-{1\over
12}H_{\mu\nu\a}H^{\mu\nu\a}=0
$$
$$
\pa_\nu(\sqrt{|g|}e^{-\phi}H^{\nu\a\b})=0\eqno(9.7)
$$
Here
$H_{\mu\nu\a}=\pa_\mu B_{\nu \a}$+ cyclic permutations is the
field strength of the antisymmetric (torsion) tensor
$B_{\mu\nu}=-B_{\nu\mu}$. I have included a general
dilaton potential, $V(\phi)$, and the possible
phenomenological contribution of other sources, represented
generically by $T_\mu^\nu$. By setting
$V=0$ and $T_{\mu\nu}=0$ we find for the
system (9.7) the particular exact (anisotropic)
solution [7] (with non-trivial torsion $H_{\mu\nu\a}\not= 0$)
$$
g_{ij}=\pmatrix{{\a+\b b^2t^2\over \b +\a b^2t^2}
 & {\sqrt{\a\b}(1+b^2t^2)\over \b +\a b^2t^2} & 0 \cr
{\sqrt{\a\b}(1+b^2t^2)\over \b +\a b^2t^2} & 1 & 0 \cr
0 & 0 & 1 \cr}
$$
$$
B_i\,^j=\pmatrix{0 & {\sqrt{\a\b}(1+b^2t^2)\over \b +\a
b^2t^2} &0 \cr
-{\sqrt{\a\b}(1+b^2t^2)\over \b +\a b^2t^2} &0 &0 \cr
0 &0 &0 \cr}
$$
$$
e^\phi =e^{\phi_0} \left[1+b^2t^2 \coth^2\left(\ga \over
2\right) \right]^{-1} , \,\,\,\,\,\,\,\,\,
\a= \cosh \ga +1 , \,\,\, \b =\cosh \ga -1. \eqno(9.8)
$$
Such solution can also be obtained by ``inverting" and
appropriately ``boosting" (through scale factor duality and
$O(2,2)$ transformations) the globally flat metric [7]
$$
ds^2=dt^2-(bt)^2dx^2-dy^2- dz^2 \eqno(9.9)
$$
($ \phi_0,\b$ and $\ga$ are free parameters).

The background (9.8) is non-trivial only in the ($x,y$) part of its
spatial sections. In order to characterize its kinematic
properties, consider the rate-of-change $H_x$ of the relative
distance along the $x$ direction, between two observers at
rest with a congruence of comoving geodesics $u^\mu$. By
projecting the expansion tensor,
$\theta_{\mu\nu}=(\nabla_\mu u_\nu +\nabla_\nu u_\mu)/2$,
on the unitary vector $n_x^\mu$ along $x$ ($n^\mu_x
u_\mu=0$,  $n_x^\mu n_{x\mu }=-1$), one easily finds
$$
H_x=-\theta_{\mu\nu}n_x^\mu n_x^\nu =
-{4t\cosh \ga \over \a \b (1+t^4) +(\a^2 +\b ^2)t^2} \eqno(9.10)
$$
(I have set $b=1$, for simplicity).

In the $t \ra -\infty$ limit
$H_x,\dot H_x, \dot \phi, \ddot \phi$ are all positive.
In the $t \ra +\infty$ limit we have instead $H_x<0$, $\dot \phi <0$,
while $\dot H_x, \ddot \phi$ are still positive.
The time evolution of $ H_x$,
(which is the analog of the Hubble parameter of
isotropic cosmological backgrounds) shows that the solution
(9.8) connects smoothly a phase of accelerated expansion of
the superinflationary type, with growing dilaton and curvature scale,
to a phase of decelerated contraction, with decreasing dilaton and
curvature scale, passing through a phase of maximal (finite) curvature.
The solution is defined over the whole time range $-\infty \leq t \leq
+\infty$,
without any singularity in the curvature and
dilaton coupling [7].

In spite of its regular behavior, this solution would seem to be of
little phenomenological interest as it is anisotropic, and the phase of
contraction (and dilaton rolling)
continues for ever down to $t=+\infty$.
Suppose, however, to perturb the above background by taking into account
the back-reaction of the produced radiation. Consider, for instance,
the amplification of
the quantum fluctuations of the metric background, whose
tensor part satisfies the equation [12,14]
$$
\ddot{h_{\omega}}  - \dot{\bar{\phi}}
 \dot{h_{\omega}} + \omega^2 h_{\omega} = 0\eqno(9.11)
$$
for each of the two physical polarization modes of proper
frequency $\om$. In the background (9.8)
$$
\fb= \phi -\ln \det |g_{\mu\nu}|^{1/2} = -\ln |bt| + const
\eqno(9.12)
$$
so that, as discussed in Sect. 3,
tensor fluctuations are amplified with a
nearly thermal spectrum, peaked around a frequency which
is of the same order as the maximal curvature scale reached
by the background.

Assuming that such scale is determined by
$\la_s$, the energy density $\r_r$ of the produced
radiation provides an initial contribution (for
$t\sim \la_s$) which is certainly subdominant, as it is
suppressed with respect to the other
terms of eq. (9.7) by the factor
$$
\left({8 \pi Ge^\phi \r_r \over |H_{\mu\nu\a}H^{\mu\nu\a}|}\right)
_{t\sim \la_s} \sim
\left(\la_p\over \la_s\right)^2 <<1 \eqno(9.13)
$$
[the expected numerical value of $\la_s$, in standard Planck units, is
given in eq.(4.3)].
However,
the radiation contribution decreases in time more
slowly than that the torsion-generated shear terms present in
eq. (9.7), as
$$
{e^\phi \r_g \over |H_{\mu\nu\a}H^{\mu\nu\a}|}\sim
|\det g_{\mu\nu}|^{-4/6}\sim t^{4/3} , \,\,\,\,\,\,\, t\ra +\infty
\eqno(9.14)
$$
The two contributions are of the same order for
$t=t_r\sim \la_s(\la_s/\la_p)^{3/2}$. For $t>>t_r$ the produced
radiation becomes then dominant and tends to isotropize the initial
solution (a well known consequence of the radiation
back-reaction [51]). In that limit the
the torsion tensor becomes negligible, and putting in eq.(9.7)
$$
g_{\mu\nu}= diag (1, -a^2(t) \da_{ij}),~~
T_\mu^\nu= diag (\r, -p\da_i^j),~~p=\r/3,~~ H_{\mu\nu\a}=0
\eqno(9.15)
$$
we can approximately describe the background evolution through the
radiation dominated, isotropic gravi-dilaton equations
$$
\dot {\fb} ^2 -dH^2 = 8 \pi G\rb e^{\fb}
$$
$$
\dot H- H\dot {\fb} ={4\pi\over 3} G\rb
e^{\fb}
$$
$$
\dot {\rb}+H \rb=0
\eqno(9.16)
$$
(we have introduced the
``shifted" variables $\fb =\phi - \ln \sqrt{|g|}$, $\rb =
\r \sqrt{|g|}$
where, in the three-dimensional isotropic case, $\sqrt{|g|}=a^3$).

By selecting the positive time branch of the general
solution [2,3] of (9.16), and imposing as initial condition a
state of decelerated contraction (to match with the previous regime of
background evolution), we are led to a particular
solution which can be written, in conformal time,
$$
a ={1\over L}\left[\eta\over \eta+\eta_0(3+\sqrt 3)
\right]^{-\sqrt 3\over 2}\left [{2\over
3}\eta^2+2\eta \eta_0(1+{1\over \sqrt 3})\right]^{1/2}
$$
$$
e^\phi=e^{\phi_0}\left[\eta
\over \eta+\eta_0(3+\sqrt 3) \right]^{-\sqrt 3} ,
\,\,\,\,\,\,\,\,\,\,\,\,\,
 \eta>0 \eqno (9.17)
$$
($ \phi_0, \eta_0, L$ are integration constants). This solution
starts from a singularity (that has been fixed, by time
translation, at $\eta=0$), and evolves from an initial
contracting, decreasing dilaton state, towards a final state of
standard radiation-dominated expansion, $a\sim t^{1/2}$, with
$\phi=$const.

If we consistently take into account the back-reaction of the
produced radiation, the background evolution may thus
approximately described by the solution (9.8) for $t<t_r$, and by
the solution (9.17) for $t>t_r$. The initial contraction is stopped and
eventually driven to expansion.
The bounce in the scale
factor, in the radiation-dominated part of the background,
marks the beginning of the standard ``post-big-bang"
regime. In this simple example the evolution
fails to be continuous at $t=t_r$,
because of the sudden approximation used to match the
torsion-driven to the radiation-driven solution. There are no
background singularities at the matching point, however, as
we are joining two different solutions within the same time
branch, $t\ra +\infty$. The dominating radiation, moreover, is entirely
produced - via quantum effects - by the classical background evolution.
\vskip 1.5 cm
\noi
{\bf 10. Conclusion}
\bigskip
\noi
In inflationary string cosmology backgrounds perturbations
can be amplified more efficiently than in conventional
inflationary backgrounds, as the perturbation amplitude my
even grow, instead of being constant, outside the horizon. In
some case, like scalar metric perturbations in a dilaton-driven
background, the effects of the growing mode can be gauged
away. But in other cases the growth is physical, and can
prevent a linearized description of perturbations.

In any case, such enhanced amplification is interesting and
worth of further study, as it may lead to phenomenological
consequences which are unexpected in the context of the
standard inflationary scenario. For instance, the production of a
relic graviton background strong enough to be detected by the
large interferometric detectors, or the
production of primordial magnetic fields strong enough to seed
the galactic dynamo. Moreover, the possible existence of a
relic stochastic electromagnetic background, due to
the amplification of the vacuum fluctuations of the
electromagnetic field, strong enough to be entirely
responsible for the observed large scale CMB anisotropy.

The main problem, in this context, is that
a rigorous and truly unambiguous discussion of all these interesting
effects would require a complete model of background evolution,
including a smooth transition from the accelerated to the decelerated
regime,through a quantum string era of Planckian curvature. A solid
string-theoretic treatment of such an
era at present is still lacking, even
if recent progress and suggestions [52] may prove useful.
The understanding of
singularities in string theory would certainly put on a firmer ground
the phenomenological
model discussed in this lecture, and might even provide a
framework for the calculation of our basic parameters $z_s, g_s$.

I have shown, nevertheless, that by including torsion in the low energy
effective action it seems possible (even to lowest order in $\ap$) to
formulate very simple models of background implementing a ``graceful
exit" from the pre-big-bang regime, at least if we accept a contracting
metric in the post-big-bang evolution. In that context a thermal
radiation background is automatically produced as a consequence
of the classical evolution, and the associate quantum back-reaction may
eventually become dominant, thus driving the background towards a final
expanding, constant dilaton regime. Moreover, thanks to the growth of
the perturbation spectrum, the identification of that radiation with the
observed CMB one is perfectly consistent with the presently
known phenomenological bounds, thus providing a framework for a unified
explanation of the $3^oK$ background and of its anisotropies.
\vskip 1.5cm
\noi
{\bf Acknowledgments.}
\bigskip
\noi
I would like to thank R. Brustein, M. Giovannini, J. Maharana, K.
Meissner, V. Mukhanov, N. Sanchez and G.
Veneziano for fruitful and enjoyable collaborations which led to
develop the ``pre-big-bang" cosmological scenario and to investigate its
possible phenomenological consequences.

\vskip 1.5 cm
\noi
{\bf 11. References}
\bigskip

\item{1.}M. Gasperini
and G. Veneziano, Astropart. Phys. 1, 317 (1993)

\item{2.} M. Gasperini and G.
Veneziano, Mod. Phys. Lett. A8, 3701 (1993)

\item{3.}M. Gasperini and G. Veneziano, Phys. Rev. D50,
2519 (1994)

\item{4.}M. Gasperini, N. Sanchez and G. Veneziano, Nucl. Phys. B364, 3
65 (1991);

Int. J. Mod. Phys. A6, 3853 (1991).

\item{5.}G. Veneziano, Phys. Lett. B265, 287 (1991)

\item{6.}K. A. Meissner and G. Veneziano, Phys. Lett. B267,
 33 (1991); Mod. Phys. Lett. A6, 3397 (1991);

M. Gasperini and G.Veneziano, Phys. Lett. B277, 256 (1992).

\item{7.}M. Gasperini, J. Maharana and G. Veneziano, Phys. Lett. B272,
277 (1991)

\item{8.}M. Gasperini, in  ``Proc. of the 2nd Journ\'ee
Cosmologie" (Observatoire de Paris, June 1994), ed. by N.
Sanchez and H. de Vega  (World Scientific, Singapore), p.429

\item{9.}G. Veneziano, {\it String cosmology: basic ideas and general
results}, these proceedings

\item{10.}L. P. Grishchuk, Sov. Phys. JEPT 40, 409 (1975);

A. A. Starobinski, JEPT Lett. 30, 682 (1979).

\item{11.}M. Gasperini and M. Giovannini, Phys. Lett. B282, 36 (1992)

\item{12.}M. Gasperini and M. Giovannini,
Phys. Rev. D47, 1529 (1992)

\item{13.}R. Brustein, M. Gasperini, M. Giovannini, V. F. Mukhanov
and G. Veneziano, Phys. Rev. D51, 6744 (1995)

\item{14.}R. Brustein, M. Gasperini, M. Giovannini and G.
Veneziano, Phys. Lett. B (1995), in press (hep-th/9507017)

\item{15.}M. Gasperini, M. Giovannini and G. Veneziano, {\it
Primordial magnetic fields from string cosmology},
CERN-TH/95-85

(hep-th/9504083)

\item{16.}M. Gasperini, M. Giovannini and G. Veneziano, {\it
Electromagnetic origin of the CMB anisotropy in string
cosmology},
CERN-TH/95-102 (astro-ph/9505041)

\item{17.}R. Brustein, M. Gasperini and M. Giovannini, {\it
Possible common origin of primordial perturbations and of the
cosmic microwave background}, Essay written for the 1995
Awards for Essays on Gravitation (Gravity Research Foundation,
Wellesley Hills, Ma), and selected for Honorable Mention
(unpublished)

\item{18.}M. Gasperini, {\it Amplification of vacuum fluctuations in
string cosmology backgrounds}, in ``Proc. of the 3rd Colloque
Cosmologie" (Observatoire de Paris, June 1995), ed. by N.
Sanchez and H. de Vega  (World Scientific, Singapore)
(hep-th/9506140)

\item{19.}M. Gasperini, M. Giovannini, K. A. Meissner and G.
Veneziano, {\it Evolution of a string network in backgrounds
with rolling horizons} (CERN-TH/95-40), to appear in
``New developments in string
gravity and physics at the Planck energy scale", ed. by N. Sanchez
(World Scientific, Singapore, 1995)

\item{20.}R. Brustein and G. Veneziano, Phys. Lett. B329,
429 (1994);

N. Kaloper, R. Madden and K. A. Olive, {\it Towards a singularity free

inflationary universe?}, UMN-TH-1333/95 (June 1995).

\item{21.}D. Shadev, Phys. Lett. B317, 155 (1984);

R. B. Abbott, S. M. Barr and S. D. Ellis, Phys. Rev. D30, 720

(1984);

E. W. Kolb, D. Lindley and D. Seckel, Phys. Rev. D30, 1205

(1984);

F. Lucchin and S. Matarrese, Phys. Lett. B164, 282 (1985).

\item{22.}A. A. Starobinski, Rel. Astr. Cosm., Byel. SSR Ac. Sci. Minsk
(1976), p.55 (in russian)

\item{23} R. B. Abbott, B. Bednarz and S. D. Ellis, Phys. Rev.
D33, 2147 (1986)

\item{24.}V. Mukhanov, H. A. Feldman and R. Brandenberger,
Phys. Rep. 215, 203 (1992)

\item{25.}L. P. Grishchuk and Y. V. Sidorov,
Phys. Rev. D42, 3413 (1990)

\item{26.}J. D. Barrow, J. P. Mimoso and M. R. de Garcia Maia,
Phys. Rev. D48, 3630 (1993)

\item{27.}S. W. Hawking, Astrophys. J. 145, 544 (1966)

\item{28.}A. R. Liddle and D. H. Lyth, Phys. Rep. 231, 1 (1993)

\item{29.}M. Bruni, G. F. R.
Ellis and P. K. S. Dunsby, Class. Quantum
Grav. 9, 921 (1992)

\item{30.}M. Gasperini, Phys. Lett. B327, 214 (1994)

\item{31.} A. Abramovici et al., Science 256, 325 (1992)

\item{32.} B. Caron et. al.,  {\it Status of the VIRGO
experiment}, Lapp-Exp-94-15

\item{33.}V. Kaplunowski, Phys. Rev. Lett. 55, 1036 (1985)

\item{34.} K. S. Thorne, in ``300 Years of Gravitation", ed. by S.
W. Hawking and W. Israel (Cambridge Univ. Press, Cambridge,
1987)

\item{35.}D. R. Stinebring et al., Phys. Rev. Lett. 65, 285 (1990)

\item{36.}V. F. Schwarztmann, JEPT Letters 9, 184 (1969)

\item{37.}P. Astone, J. A. Lobo and B. F. Schutz, Class.
Quantum Grav. 1, 2093 (1994)

\item{38.} F. Pegoraro, E. Picasso, L. Radicati, J. Phys.
A1, 1949 (1978);

C. M. Caves, Phys. Lett. B80, 323 (1979);

C. E. Reece et al., Phys. Lett A104, 341 (1984).

\item{39.} M. S. Turner and L. M. Widrow,
Phys. Rev. D37, 2743 (1988)

\item{40.} B. Ratra, Astrophys. J. Lett. 391, L1 (1992);

A. D. Dolgov, Phys. Rev. D48, 2499, (1993).

\item{41.}D. Lemoine and M. Lemoine, {\it Primordial magnetic
fields  in string cosmology}, April 1995

\item{42.} E. N. Parker,
``Cosmical Magnetic fields" (Clarendon, Oxford,
England, 1979)

\item{43.}G. F. Smoot et al., Astrophys. J. 396, L1 (1992)

\item{44.}C. L. Bennett et al., Astrophys. J. 430, 423 (1994)

\item{45.}D. Grasso and H. R. Rubinstein, Astropart. Phys. 3, 95
(1995)

\item{46.}B. Zwiebach, Phys. Lett. B156, 315 (1985)

\item{47.}E. W. Kolb and M. S. Turner, {\it The Early Universe}
(Addison-Wesley, Redwood City, Ca, 1990)

\item{48.}L. Parker, Nature 261, 20 (1976)

\item{49.}N. D. Birrel and P. C. W. Davies, {\it Quantum fields in
curved spaces} (Cambridge Univ. Press, Cambridge, England, 1982);

B. Allen, Phys. Rev. D37 (1988) 2078;

J. Garriga and E. Verdaguer, Phys. Rev. D39 (1989) 1072;

C. R. Stephens, Phys. Lett. A142 (1989) 68.

\item{50.}C. Lovelace, Phys. Lett. B135 (1984) 75;

E. S. Fradkin and A. A. Tseytlin, Nucl. Phys. B261 (1985) 1;

C. G. Callan et al., Nucl. Phys. B262 (1985) 593.

\item{51.}J. B. Zeldovich and I. D. Novikov, {\it Relativistic
astrophysics},

(Chicago University Press, 1983)

\item{52.}E. Kiritsis and K. Kounnas, Phys. Lett. B331, 51 (1994);

A. A. Tseytlin, Phys. Lett. B334, 315 (1994);

E. Martinec, Class. Quantum Grav. 12, 941 (1995).

\end